\documentclass[onecolumn,secnumarabic,amssymb, nobibnotes, aps, prd]{revtex4}
\usepackage{amsmath} \usepackage{amssymb}
\usepackage{graphicx} 
\usepackage{epstopdf}
\usepackage{amsmath}
\usepackage{amsmath,amssymb,amsthm,amsfonts,mathrsfs,bm,verbatim}
\usepackage{graphicx,subfigure}

\newcommand{\bea}{\begin{eqnarray}}
\newcommand{\eea}{\end{eqnarray}}
\def\nn{\nonumber}

\begin{document}

\title{The superradiant stability of Kerr-Newman black holes }%

\author{Wen-Xiang Chen$^{a}$}
\affiliation{Department of Astronomy, School of Physics and Materials Science, GuangZhou University, Guangzhou 510006, China}
\author{Yao-Guang Zheng}
\email{hesoyam12456@163.com}
\affiliation{Department of Astronomy, School of Physics and Materials Science, GuangZhou University, Guangzhou 510006, China}

\begin{abstract}
In this article, the superradiation stability of Kerr-Newman black holes is discussed by introducing the condition used in Kerr black holes y into them. Moreover, the motion equation of the minimally coupled scalar perturbation in a Kerr-Newman black hole is divided into angular and radial parts. Hod proved\cite{12} that the Kerr black hole should be superradiantly stable under massive scalar perturbation when $\mu \ge \sqrt{2}m\Omega_H$, where $\mu$ is the mass. In this article, a new variable y is added here to expand the results of the above article. When $\sqrt{2(a^2+Q^2)}/{r^2_+}< \omega< m\varOmega_H+q\varPhi_H$,particularly $\mu \ge \sqrt{2}(m\varOmega_H+q\varPhi_H)$,so the Kerr-Newman black hole is superradiantly stable at that time.

  \textbf{Keywords: superradiantly stable, a new variable y, Kerr-Newman black holes}

\end{abstract}

\maketitle

\section{Introduction}
 In the 1970s, the development of black hole thermodynamics applies the basic laws of thermodynamics to the theory of black holes in the field of general relativity and strongly implies the profound and basic relationship between general relativity, thermodynamics, and quantum theory. 
The stability of black holes is a major topic in black hole physics. Regge and Wheeler\cite{2} have proved that the spherically symmetric Schwarzschild black hole is stable under perturbation. The great impact of superradiance makes the stability of rotating black holes more complicated. Superradiative effects occur in both classical and quantum scattering processes. Condition (1) characterizes super-radiantly amplified scalar modes that could lead to instabilities of the central (charged and spinning) Kerr-Newman black holes.
\bea\label{src}
\omega< m\varOmega_H+q\varPhi_H,  \varOmega_H=\frac{a}{r_+^2+a^2}
\eea
where $q$ and $m $ are the charge and azimuthal quantum number of the incoming wave, $\omega$ denotes the wave frequency, $\varOmega_H$ is the angular velocity of the black hole horizon and $\varPhi_H$ is the electromagnetic potential of the black hole horizon. If the frequency range of the wave lies in the superradiance condition, the wave reflected by the event horizon will be amplified, which means the wave extracts rotational energy from the rotating black hole when the incident wave is scattered. According to the black hole bomb mechanism proposed by Press and Teukolsky\cite{1,2,3,4,5,6,7,8,9}, if a mirror is placed between the event horizon and the outer space of the black hole, the amplified wave will reflect back and forth between the mirror and the black hole and grow exponentially, which leads to the super-radiative instability of the black hole.

Superradiant refers to the enhancement process of emission, flow density, and intensity when a certain frequency wave or current density and intensity pass through the medium or the edge of the medium. In quantum mechanics, the superradiant effect can be traced back to Klein's fallacy. Subsequently, more superradiation phenomena are discovered in classical physics and quantum mechanics. In 1971\cite{1,2,3,4,5,6,7,8,9}, Zel'dovich proved that a rotating conducting cylinder magnifies the scattered waves. In 1972\cite{3,4,5,6,7,8,9}, Misner first established a quantitative theory of the superradiance effect of black holes. Press and Teukolsky proposed a black hole bomb mechanism due to the existence of superradiation mode. If there is a mirror between the event horizon and the space of the black hole, the amplified wave can scatter back and forth and grow exponentially, which leads to the superradiation instability in the black hole background. Specifically, the black hole bomb is the name of the physical effect. It is a physical phenomenon produced by the impact of a boson field amplified by superradiation scattering on a rotating black hole. The additional condition that the phenomenon must satisfy is that the field must have a static mass that is not equal to zero. In this phenomenon, the scattered wave will be reflected back and forth between the mass interference term and the black hole and will be amplified at each reflection.

The superradiation stability of black holes has always been an open research topic. The problems of radiation stability and steady-state resonance have been discussed by many authors in recent years, such as S. Hod, Ran Li, V. Cardoso, AN Aliev, etc. It has been extensively and in-depth studied using analytical and numerical methods.S. Hod and C. Herdeiro et al\cite{40,41} used analytical and numerical methods to systematically study the steady-state resonance of the Kerr-Newman black hole.

Although there has been so much study on superradiance of rotating black holes, even Kerr black hole is not investigated thoroughly. Hod proved\cite{12} that the Kerr black hole should be
superradiantly stable under massive scalar perturbation when $\mu \ge \sqrt{2}m\Omega_H$, where $\mu$ is the mass. 

In this article, a new variable y is added here to expand the results of the above article. When $\mu \ge \sqrt{2}(m\varOmega_H+q\varPhi_H)$,so\ the Kerr-Newman black\ hole\ is\ superradiantly\ stable\ at\ that\ time. (1)We give an outline of the  Kerr-Newman-black-hole-massive-scalar system and the angular part of the equation of motion in Section 2. (2)We discuss some important asymptotic behaviors of the effective potential and its derivatives based on the Schrodinger-like radial equation and the effective potential of scalar perturbation in Kerr-Newman  background in Section 3. (3)We obtain the parameter space region of superradiation stability by analyzing the effective potential in detail in Section 4. (4)We get the limit $y$ of the incident particle under the superradiance of Kerr-Newman black holes in Section 5. (5) We get their relation to thermodynamic geometry and superradiant stability in Section 6.

\section{The system of Kerr-Newman black hole}

 The metric of the 4-dimensional Kerr-Newman black hole under Boyer-Lindquist coordinates $(t,r,\theta,\phi)$ is written as follow (with natural unit, $G=\hbar=c=1$)\cite{11,12,13,14,15,16,17,18,19,20,21,22,23,24}
\bea\nn
ds^2&=&-\frac{\varDelta}{\rho ^2}\left( dt-a \sin ^2\theta d\phi \right) ^2+\frac{\rho ^2}{\varDelta}dr^2\\
&+&\rho ^2d\theta ^2+\frac{\sin ^2\theta}{\rho ^2}\left[ \left( r^2+a^2 \right) d\phi -at \right] ^2,
\eea
where
\begin{equation}
\varDelta\equiv r^2-2Mr+a^2+Q^2\quad,\quad\rho ^2\equiv r^2+a^2\cos ^2\theta,
\end{equation}
$a$ denotes the angular momentum per unit mass of a certain  Kerr-Newman black hole and $Q,~M$ denotes its charge and mass.
The inner and outer horizons of the Kerr-Newman black hole can be expressed as
\begin{equation}
{{r}_{\pm }}=M\pm \sqrt{{{M}^{2}}-{{a}^{2}}-{{Q}^{2}}},
\end{equation}
and obviously 
\begin{equation}
{{r}_{+}}+{{r}_{-}}=2M,~~{{r}_{+}}{{r}_{-}}={{a}^{2}}+{{Q}^{2}}.
\end{equation}
The background electromagnetic potential is written as follows
\begin{equation}
A_{\nu}=\left( -\frac{Qr}{\rho ^2},0,0,\frac{aQr\sin ^2\theta}{\rho ^2} \right).
\end{equation}

 The following covariant Klein-Gordon equation
\begin{equation}
( \nabla ^{\nu}-iqA^{\nu})( \nabla _{\nu}-iqA_{\nu}) \Phi =\mu ^2\Phi,
\end{equation}
where $\nabla ^{\nu}$ represents the covariant derivative under the Kerr-Newman background. We adopt the method of separation of variables to solve the above equation, and it is decomposed as
\begin{equation}
\Phi ( t,r,\theta ,\phi) =\sum_{lm}R_{lm}( r ) S_{lm}( \theta) e^{im\phi}e^{-i\omega t}.
\end{equation}
where $R_{lm}$ are the equations that satisfy the radial equation of motion (see Eq.\eqref{REoM} below). The angular function $S_{lm}$ denotes the scalar spheroidal harmonics which satisfy the angular part of the equation of motion (see Eq.\eqref{AEoM} below). $l(=0,1,2,...)$ and $m$ are integers, $-l\leq m\leq l$ and $\omega$ denote the angular frequency of the scalar perturbation. 

The angular part of the equation of motion is an ordinary differential equation and it can be expressed as follows, 
\bea\label{AEoM}\nn
&&\frac{1}{\sin \theta}\frac{d}{d\theta}( \sin \theta \frac{dS_{lm}}{d\theta})\\
 &&+[ K_{lm}+( \mu ^2-\omega ^2) a^2\sin ^2\theta -\frac{m^2}{\sin ^2\theta} ] S_{lm}=0,
\eea
where $K_{lm}$ represent angular eigenvalues. The standard spheroidal differential equation above has been studied for a long time and is of great significance in plenty of physical problems. The spheroidal functions $S_{lm}$ are known as prolate (oblate) for $( \mu ^2-\omega ^2) a^2>0( <0)$, and only the prolate case is concerned in this article. 
We select the lower bound for this separation constant as follow,\cite{25,26,27,28,29,30,31,32,33,34,35}
\begin{equation}
K _{lm}\geq m^2 -a^2( \mu ^2-\omega ^2).
\end{equation}

The radial part of the Klein-Gordon equation contented by $R_{lm}$ is written as
\begin{equation}\label{REoM}
\varDelta \frac{d}{dr}( \varDelta \frac{dR_{lm}}{dr}) +UR_{lm}=0,
\end{equation}
where
\bea\nn
U&=&[\omega( a^2+r^2)-am-qQr] ^2\\
&&+\varDelta [ 2am\omega-K _{lm}-\mu ^2( r^2+a^2)].
\eea

The study of the superradiant modes of KN black hole under the charged massive scalar perturbation requires considering the asymptotic solutions of the radial equation near the horizon and at spatial infinity for appropriate boundary conditions. Here we adopt a tortoise coordinate to analyze the boundary conditions for the radial function. 
The tortoise coordinate $r_*$ is defined by the following equation
\begin{equation}
\frac{dr_*}{dr}=\frac{r^2+a^2}{\varDelta}.
\end{equation}

The two boundary conditions we focused on are the purely ingoing wave next to the outer horizon and the exponentially decaying wave located at spatial infinity. Thus the asymptotic solutions of the radial wave function under the above boundaries are selected as follows
\bea\nn
R_{lm}( r) \sim \begin{cases}
	e^{-i( \omega -\omega _c) r_*}, &r^*\rightarrow -\infty\text{(}r\rightarrow r_+ \text{)}\\
	\frac{e^{-\sqrt{\mu ^2-\omega ^2}r}}{r}, &r^*\rightarrow +\infty\text{(}r\rightarrow +\infty \text{)}.\\
\end{cases}
\eea
We can easily see that getting decaying modes at spatial infinity requires following bound state condition
\begin{equation}\label{bsc}
\omega ^2<\mu^2.
\end{equation}
The critical frequency $\omega_c$ is defined as
\begin{equation}
\omega _c=m\varOmega _H+q\varPhi _H,
\end{equation}
where $\varOmega _H$ is the angular velocity of the outer horizon and $\varPhi _H=\frac{Qr_+}{r_{+}^{2}+a^2}$ is the electric potential of whom.

\section{The radial equation of motion and effective potential}

A new radial wave function is defined as\cite{11,30,31,32,33,34,35}
\begin{equation}
\psi _{lm}\equiv \varDelta^{\frac{1}{2}}R_{lm}.
\end{equation}
to substitute the radial equation of motion \eqref{REoM} for a Schrodinger-like equation
\begin{equation}
\frac{d^2\Psi _{lm}}{dr^2}+( \omega ^2-V) \Psi _{lm}=0,
\end{equation}
where
\begin{equation}
\omega ^2-V=\frac{U+M^2-a^2-Q^2}{\varDelta ^2},
\end{equation}
in which $V$ denotes the effective potential. 

Taking the superradiant condition \eqref{src}, i.e. $\omega<\omega_c$, and bound state condition \eqref{bsc} into consideration,  the  Kerr-Newman black hole and charged massive scalar perturbation system are superradiantly stable when the trapping potential well outside the outer horizon of the  Kerr-Newman black hole does not exist. As a result, the shape of the effective potential $V$ is analyzed next to inquiry into the nonexistence of trapping well.

The asymptotic behaviors of the effective potential $V$ around the inner and outer horizons and at spatial infinity can be expressed as
\begin{equation}
V( r\rightarrow +\infty )
\rightarrow \mu ^2-\frac{2( 2M\omega ^2-qQ\omega -M\mu ^2)}{r}+{\cal O}( \frac{1}{r^2}) ,
\end{equation}
\bea
V( r\rightarrow r_+ ) \rightarrow -\infty,~~
V( r\rightarrow r_-) \rightarrow -\infty.
\eea

If a Kerr black hole satisfies the condition of $\mu=y\omega$, it will be superradiantly stable when $\mu<\sqrt{2}m\Omega_H$. In this article, we introduce the above condition to Kerr-Newman black holes. Therefore, the formula of the asymptotic behaviors is written as

\begin{equation}
V( r\rightarrow +\infty )
\rightarrow y^2\omega^2-\frac{2[M(2-y^2)\omega ^2-qQ\omega] }{r}+{\cal O}( \frac{1}{r^2}) ,
\end{equation}
\bea
V( r\rightarrow r_+ ) \rightarrow -\infty,~~
V( r\rightarrow r_-) \rightarrow -\infty.
\eea
It is concluded from the equations above that the effective potential approximates a constant at infinity in space, and the extreme between its inner and outer horizons cannot be less than one. The asymptotic behavior of the derivative of the effective potential $V$ at spatial infinity can be expressed as 
\begin{equation}
 V'( r\rightarrow +\infty )
 \rightarrow \frac{2[ M(2-y^2)\omega ^2-qQ\omega ]}{r^2}+{\cal O}( \frac{1}{r^3}) ,
\end{equation}
The derivative of the effective potential has to be negative to satisfy the no trapping well condition,
\begin{equation}
2M(2-y^2)\omega^2-2Qq\omega<0.
\end{equation}

\section{Algebraic analysis of the superradiant stability of Kerr-Newman black holes}
 A new radial coordinate is defined as $z$, $z=r-{r_-}$. The explicit expression of the derivative of the effective potential $V$ in radial coordinates $z$ and $r$ is expressed as\cite{37,38,39} 
\bea\nonumber
V'( r)&=&\frac{Ar^4+Br^3+Cr^2+Dr+E}{-\varDelta ^3}\\
=V'( z)&=&\frac{A_1z^4+B_1z^3+C_1z^2+D_1z+E_1}{-\varDelta ^3}=\frac{f_1(z)}{-\varDelta ^3}.
\eea
The two sets of coefficients satisfy the following relation, where
\bea
A_1&=& A,\\
{B_1} &=& B+4r_-A_1,\\
{C_1} &=& C+3r_-B_1-6r_{-}^{2}A_1,\\
{D_1} &=& D+4r_{-}^{2}A_1-3r_{-}^{2}B_1+2r_-C_1,\\
{E_1} &=& E-r_{-}^{4}A_1+r_{-}^{3}B_1-r_{-}^{2}C_1+r_-D_1.
\eea

The complete expressions of coefficients $A_1$, $E_1$ and $C_1$ are as follows

\bea
A_1&=&2qQ\omega +2M\mu ^2-4M\omega ^2,\\\nonumber
E_1&=&-4a^2M+4a^2m^2M+4M^3-4MQ^2\\\nn
&& +2amqQ( a^2+Q^2) +4a^2M( -a^2-Q^2) \mu ^2\\\nn
&& +2a^2M( a^2+Q^2) \mu ^2+4M( -a^2-Q^2) K_{lm}\\\nn
&& +2M( a^2+Q^2) K_{lm}+4a^2r_--4a^2m^2r_-\\\nn
&& +4amMqQr_-+2a^2q^2Q^2r_-+4a^2M^2\mu ^2r_-\\\nn
&& -2a^2Q^2\mu ^2r_-+4M^2( -1+K_{lm}) r_-\\\nn
&& +2a^2K_{lm}r_-+2Q^2[ 2+Q^2( q-\mu)( q+\mu) \\\nn
&& +K_{lm}] r_--6amqQr_{-}^{2}-6M( -Q^2\mu ^2\\\nn
&& +K_{lm}) r_{-}^{2}-4M^2\mu ^2r_{-}^{3}-2Q^2( q^2+\mu ^2) r_{-}^{3}\\\nn
&& +2K_{lm}r_{-}^{3}+2M\mu ^2r_{-}^{4}+\omega ^2( 4a^4M+4a^2Q^2r_-\\\nn
&& +4Q^2r_{-}^{3}-4Mr_{-}^{4}) +\omega[ -8a^3mM\\\nn
&& -8amM( -a^2-Q^2) -4amM( a^2+Q^2) \\\nn
&& -2a^2qQ( a^2+Q^2)-4a^2MqQr_-\\\nn
&& -8am( M^2+Q^2) r_-+12amMr_{-}^{2}-6qQ^3r_{-}^{2}\\
&& +4MqQr_{-}^{3}+2qQr_{-}^{4}].
\eea
\bea\nn
C_1&=&-3(r_+-r_-)K_{lm}+12r_-(Q^2-r_-^2-r_+r_-)\omega^2\\\nonumber
&&+6[am(r_-+r_+)-(Q^2-3r_-^2-r_-r_+)qQ]\omega\\\nonumber
&&-3(Q^2r_--Q^2r_+-r_-^3+r_+^2 r_-)\mu^2\\
&&-6amqQ-6q^2Q^2r_-,
\eea

In this article, we represent the numerator of the derivative of the effective potential $V'(z)$. This quartic polynomial of $z$ enables us to study the existence of the trapping well outside the horizon by analyzing the property of the roots of the equation. We adopt $z_1$, $z_2$, $z_3$ and $z_4$ to denote the four roots of $f_1(z) = 0$. The relationships among them adhere to the Vieta theorem. 
\bea
z_1z_2z_3z_4=\frac{E_1}{A_1},\\
z_1z_2+z_1z_3+z_1z_4+z_2z_3+z_2z_4+z_3z_4=\frac{C_1}{A_1}.
\eea

When $z> 0$, one can infer from the asymptotic behaviors of the effective potential at the inner and outer horizons and spatial infinity that the number of positive roots of $V'(z)=0(\text{or}~ f_1(z)=0)$ cannot less than two. Hence these two positive roots are written as $z_1, z_2$.

Studies have shown that for any $\omega$
\bea
E_1>0.
\eea  
and under the condition of 
\bea
E_1>0,~~C_1<0,
\eea 
$f_1(z)=0$, i.e., $z_3,z_4$ are both negative.  

The expression of $C_1$ is written as
\bea\label{C1111}\nonumber
\tiny C_1&=&-3(r_+-r_-)K_{lm}+12r_-(Q^2-r_-^2-r_+r_-)\omega^2\\\nonumber
&&+6[am(r_-+r_+)-(Q^2-3r_-^2-r_-r_+)qQ]\omega\\\nonumber
&&-3(Q^2r_--Q^2r_+-r_-^3+r_+^2 r_-)\mu^2\\
&&-6amqQ-6q^2Q^2r_-.
\eea
Here, the eigenvalue of the spheroidal angular equation $K _{lm}$ follows the lower bound,
\bea
K _{lm}\geq m^2 -a^2( \mu ^2-\omega ^2).
\eea

When $\mu=y\omega$, under the consideration of the inequality \eqref{bsc}, the further expression of the above inequality is
\begin{equation}
\begin{aligned}
C_1<&3[2(a^2+2Q^2)r_-^2-(a^2+4r_-^2)2M-\\
&r_-^2y^2(r_++r_-)]\omega^2\\
&+6[-qQ^3+qr_-(2r_-+2M)Q+2amM]\omega\\
&-6q^2r_-Q^2-6amqQ-6m^2\sqrt{M^2-a^2-Q^2}.
\end{aligned}
\end{equation}
The right side of the above inequality is denoted as $f(\omega)$ $i.e$ a quadratic function with $\omega$ as independent variable
\begin{equation}
\begin{aligned}
f(\omega)=&3[2(a^2+2Q^2)r_-^2-(a^2+4r_-^2)2M-\\
&r_-^2y^2(r_++r_-)]\omega^2\\
&+6[-qQ^3+qr_-(2r_-+2M)Q+2amM]\omega\\
&-6q^2r_-Q^2-6amqQ-6m^2\sqrt{M^2-a^2-Q^2}.
\end{aligned}
\end{equation}
Let 
\begin{equation}
    a1=3[2(a^2+2Q^2)r_-^2-(a^2+4r_-^2)2M-r_-^2y^2(r_++r_-)]
\end{equation}
\begin{equation}
    b1=6[-qQ^3+qr_-(2r_-+2M)Q+2amM]
\end{equation}
\begin{equation}
   c1=-6q^2r_-Q^2-6amqQ-6m^2\sqrt{M^2-a^2-Q^2}
\end{equation}
In the case of $f(\omega)<0$, the numerical value of $m$ is a problem. We next discuss the $b1\omega+c1$ part of $f(\omega)$. When $b1\omega+c1<0$, $a1<0$ both conditions are satisfied, $f(\omega)<0$. $b1\omega+c1<0$ can be regarded as a problem of $M$ quadratic equation less than 0. As a result, we suppose when $\omega<\frac{qQ}{2M}$, $b1\omega+c1<0$. We obtain a sufficient condition of $f(w)<0$ from the above equation
\begin{equation}
    H(m)=-\sqrt{M^2-a^2-Q^2}m^2
    +a(-qQ+2M\omega)m-qQ(qQr_-^2+(Q^2-2r_-(M+r_-))\omega)
\end{equation}
The quadratic equation of $b1\omega+c1$, when 
\begin{equation}
    m > -\frac{-a(-qQ+2M\omega)}{2\sqrt{-a^2+M^2-Q^2}}
    -\frac{a^2(-qQ+2M\omega)^2-4qQ\sqrt{-a^2+M^2-Q^2}(qQr_-^2+Q^2-2r_-(M+r_-))\omega}{2\sqrt{-a^2+M^2-Q^2}}
\end{equation}
As we can see that $b1\omega+c1<0$ when $\omega<\frac{qQ}{2M}$. 

In the parameter region where $f(\omega)<0$ is a sufficient condition for $C_1<0$. Firstly, the quadratic coefficient of $f(\omega)$ is negative. Therefore, the following inequality is derived

\begin{equation}
y^2(r_+-r_-)\frac{r_-}{r_+}>2r_-+\frac{2Q^2}{r_+}-a^2-4r_-^2
\end{equation}

\begin{equation}
r_+>\frac{2(a^2+2Q^2)}{a^2+4r_-^2}
-\frac{2\sqrt{M^2-a^2-Q^2}(M-\sqrt{M^2-a^2-Q^2})y^2}{a^2+4r_-^2}.
\end{equation}
 Eq. (48) is deduced from the relation $f(\omega)<0$.

When ${y}^2 >2({A_1}>0)$\,
for\ ${E_1}>0$, $ {C_1}<0$\ at\ this\ time,\ then $ f(\omega) < 0$,\ and\ we\ can\ know\ that\ the\ equation\ $ \text{ }{{\text{V}}_{\text{1}}}\text{ }\!\!'\!\!\text{ (z)}=\text{0}$\ cannot\ have\ more\ than\ two\ positive\ roots.\ So\ the Kerr-Newman black\ hole\ is\ superradiantly\ stable\ at\ that\ time.

\section{The limit $y$ of the incident particle under the superradiance of Kerr-Newman black holes}
We will study the physical and mathematical properties of the linearized large mass scalar field configuration (scalar cloud) with nontrivial coupling to the electromagnetic field of the Kerr-Newman black hole. The line element of space-time of spherically symmetric Kerr-Newman black hole can be expressed in the form of \cite{41}

\bea\nn
ds^2=-g(r)dt^2+({1/{g(r)}})dr^2+r^2(d\theta^2+\sin^2\theta
d\phi^2),
\eea
where
\begin{equation}\label{Eq2}
g(r)=1-{{2M}/{r}}+({{a^2+Q^2})/{r^2}}\  .
\end{equation}

V can change to \cite{17}
\begin{equation}
V(r)=\left(1-\frac{2 M}{r}+\frac{({a}^{2}+{Q}^{2})}{r^{2}}\right)\left[\mu^{2}+\frac{l(l+1)}{r^{2}}+\frac{2 M}{r^{3}}-\frac{2 ({a}^{2}+{Q}^{2})}{r^{4}}-\frac{\alpha ({a}^{2}+{Q}^{2})}{r^{4}}\right]
\end{equation}

As we will show clearly now, the Schrodinger-like equation determines the radial function behavior of the space-bounded non-minimum coupled mass scalar field configuration of Kerr-Newman black hole space-time and is suitable for WKB analysis of large mass systems. In particular, Schrodinger-like standard second-order WKB analysis of the radial equation produces the well-known discrete quantization condition(Here we have used the integral relation $\int_{0}^{1} d x \sqrt{1 / x-1}=\pi / 2$.),when $V( r\rightarrow +\infty ),\mu=1/(n+{1\over2})$, 
\begin{equation}
\int_{(y^2)_{t-}}^{(y^2)_{t+}}d(y^2)\sqrt{\omega ^2-V1(y;M,a,l,\mu,\alpha1)}=\big(n+{1\over2}\big)\cdot\pi\mu /{{2}}=\pi/2
\ \ \ ; \ \ \ \ n=0,1,2,...\  .
\end{equation}
The two integration boundaries $\{y_{t-},y_{t+}\}$ of the WKB
formula is the classical turning points [with
$V(y_{t-})=V(y_{t+})=0$] of the composed
charged-black-hole-massive-field binding potential.
The resonant parameter $n$ (with $n\in\{0,1,2,...\}$) characterizes
the infinitely large discrete resonant spectrum
$\{\alpha_n(\mu,l,M,a)\}_{n=0}^{n=\infty}$ of the black-hole-field
system.

Using the relation  between the radial coordinates $y$ and $r$, one can
express the WKB resonance equation  in the form
\begin{equation}
\int_{r_{t-}}^{r_{t+}}dr{{\sqrt{-V(r;M,a,l,\mu,\alpha)}}\over{g(r)}}=\big(n+{1\over2}\big)\cdot\pi\
\ \ \ ; \ \ \ \ n=0,1,2,...\  ,
\end{equation}
where the two polynomial relations 
\begin{equation}
1-{{2M}\over{r_{t-}}}+{{a^2+Q^2}\over{r^2_{t-}}}=0\
\end{equation}
and
\begin{equation}
{{l(l+1)}\over{r^2_{t+}}}+{{2M}\over{r^3_{t+}}}-{{2(a^2+Q^2)}\over{r^4_{t+}}}
-{{\alpha (a^2+Q^2)}\over{r^4_{t+}}}=0\
\end{equation}
determine the radial turning points $\{r_{t-},r_{t+}\}$ of the composed black-hole-field binding potential.

We set
\begin{equation}
x\equiv {{r-r_{\text{+}}}\over{r_{\text{+}}}}\ \ \ \ ; \ \ \ \ \tau\equiv {{r_+-r_-}\over{r_+}}\  ,
\end{equation}
in terms of which the composed black-hole-massive-field interaction term 
has the form of a binding potential well,
\begin{equation}
V[x(r)]=-\tau\Big({{\alpha(a^2+Q^2)}\over{r^4_+}}-\mu^2\Big)\cdot x +
\Big[{{\alpha(a^2+Q^2)(5r_+-6r_-)}\over{r^5_+}}-\mu^2\big(1-{{2r_-}\over{r_+}}\big)\Big]\cdot x^2+O(x^3)\  ,
\end{equation}
in the near-horizon region
\begin{equation}
x\ll\tau\  .
\end{equation}

From the near-horizon expression  of the
black-hole-field binding potential, one obtains the dimensionless
expressions
\begin{equation}
x_{t-}=0\
\end{equation}
and
\begin{equation}
x_{t+}=\tau\cdot{{{{\alpha (a^2+Q^2)}\over{r^4_+}}-\mu^2}\over{{{\alpha (a^2+Q^2)(5r_+-6r_-)}\over{r^5_+}}-\mu^2\big(1-{{2r_-}\over{r_+}}\big)}}\
\end{equation}
for the classical turning points of the WKB integral relation.

We find that
our analysis is valid in the regime  below($\alpha1$ corresponds to the transformation of y )
\begin{equation}
\alpha\simeq{{\mu^2r^4_+}\over{(a^2+Q^2)}}  ,\alpha1\simeq\sqrt{{{\mu^2r^4_+}\over{(a^2+Q^2)}}}\
\end{equation}
in which case the near-horizon binding potential and its outer
the turning point can be approximated by the remarkably compact
expressions
\begin{equation}
V(x)=-\tau\Big[\Big({{\alpha (a^2+Q^2)}\over{r^4_+}}-\mu^2\Big)\cdot
x-4\mu^2\cdot x^2\Big]+O(x^3)\
\end{equation}
and
\begin{equation}
x_{t+}={1\over4}\Big({{\alpha (a^2+Q^2)}\over{\mu^2r^4_+}}-1\Big)\  .
\end{equation}
In addition, one finds the near-horizon relation
\begin{equation}
p(x)=\tau\cdot x+(1-2\tau)\cdot x^2+O(x^3)\  .
\end{equation}

We know that
\begin{equation}
{{1}\over{\sqrt{\tau}}}\int_{0}^{x_{t+}}dx \sqrt{{{{\alpha
(a^2+Q^2)}\over{r^2_+}}-\mu^2
r^2_+\over{x}}-4\mu^2r^2_+}=\big(n+{1\over2})\cdot\pi\ \ \ \ ; \ \ \
\ n=0,1,2,...\  .
\end{equation}
Defining the dimensionless radial coordinate
\begin{equation}
z\equiv {{x}\over{x_{t+}}}\  ,
\end{equation}
one can express the WKB resonance equation the mathematically
compact form
\begin{equation}
{{2\mu r_+ x_{t+}}\over{\sqrt{\tau}}}\int_{0}^{1}dz
\sqrt{{{1}\over{z}}-1}=\big(n+{1\over2})\cdot\pi\ \ \ \ ; \ \ \ \
n=0,1,2,...\  ,
\end{equation}
which yields the relation 
\begin{equation}
{{\mu r_+ x_{t+}}\over{\sqrt{\tau}}}=n+{1\over2}\ \ \ \ ; \ \ \ \
n=0,1,2,...
\end{equation}

We know from the curve integral formula that there is a certain extreme value forming a loop
\begin{equation}
1/y^2 \rightarrow {\alpha}
\end{equation}y takes the interval from 0 to 1 at this time.

 The physical parameter
$y$ is defined by the dimensionless relation, for $y$ is greater than $\sqrt{2}$ at this time,
\begin{equation}\label{Eq33}
{y}\equiv{\alpha1}/\sqrt{2} .
\end{equation} 

Here the critical parameter y is given by the
simple relation 
\begin{equation}
{y}/\mu\equiv {{r^2_+}\over\sqrt{2(a^2+Q^2)}}  .
\end{equation}
When
\bea
\sqrt{2(a^2+Q^2)}/{r^2_+}< \omega< m\varOmega_H+q\varPhi_H,  \varOmega_H=\frac{a}{r_+^2+a^2}
\eea
the Kerr-Newman black\ hole\ is\ superradiantly\ stable\ at\ that\ time.

 \section{The thermodynamic geometry of Kerr-Newman black holes and their relation to superradiant stability}
We will show that Hawking radiation from Kerr-Newman black holes can be understood as fluxes that counteract gravitational anomalies. The point is that near the event horizon, the scalar field theory in the spacetime of a 4-Vick black hole can be reduced to a two-dimensional field theory. Since spacetime is not spherically symmetric, this is an unexpected result.\cite{42}

We can rewrite the action
 \begin{equation}
S[\varphi]=\frac{a}{2 \Omega_{H}} \int d^{4} x \sin \theta \varphi\left(-\frac{1}{f(r)} \partial_{\xi}^{2}+\partial_{r} f(r) \partial_{r}\right) \varphi
\end{equation}
We know that when $\sin \theta$ = 0, the pull equation for action can conform to the above form, but the boundary becomes 0. The action form can always be reduced to the form of negative power expansion. It can be seen that Hawking radiation is consistent with superradiation.

Black hole thermodynamics or black hole mechanics is a theory developed in the 1970s by applying the fundamental laws of thermodynamics to the study of black holes in the field of general relativity. Although the theory is not yet clearly understood, the existence of black hole thermodynamics strongly suggests a deep and fundamental connection between general relativity, thermodynamics, and quantum theory. Although it seems to only start from the most basic thermodynamic principles and describe the behavior of black holes under the constraints of the laws of thermodynamics through classical and semi-classical theories, its significance goes far beyond the analogy between classical thermodynamics and black holes. Related to the nature of quantum phenomena in gravitational fields.

Thermodynamic geometry of Kerr-Newman black holes:Entroy S\cite{43}
\begin{equation}
 S\rightarrow4 \pi\left[2 M\left(M+\sqrt{M^{2}-a^{2}-Q^{2}}\right)-Q^{2}\right]
\end{equation}

Through the thermodynamic geometric metric, we obtain the expression of the Kerr-Newman black hole Ruppeiner metric
\begin{equation}
g_{a b}^{R}=\frac{\partial^{2}}{\partial x^{a} \partial x^{b}} S(M, Q) \quad(a, b=1 ,2)
\end{equation}
where $x^{1}=H, x^{2}=\Omega$. By calculation, we get the metric expression
\begin{equation}
g_{1 1}^{R}=8 \pi\left(2-\frac{M^{3}}{\left(-a^{2}+M^{2}-Q^{2}\right)^{3 / 2}}+\frac{3 M}{\sqrt{-a^{2}+M^{2}-Q^{2}}}\right)
\end{equation}
\begin{equation}
g_{1 2}^{R}=\frac{8 \pi Q\left(a^{2}+Q^{2}\right)}{\left(-a^{2}+M^{2}-Q^{2}\right)^{3 / 2}}
\end{equation}
\begin{equation}
g_{2 1}^{R}=
\frac{8 \pi Q\left(a^{2}+Q^{2}\right)}{\left(-a^{2}+M^{2}-Q^{2}\right)^{3 / 2}} 
\end{equation}
\begin{equation}
g_{2 2}^{R} =8 \pi\left(-1+\frac{M\left(a^{2}-M^{2}\right)}{\left(-a^{2}+M^{2}-Q^{2}\right)^{3 / 2}}\right)
\end{equation}

The curvature scalar of a Kerr-Newman black hole is
\begin{equation}
\begin{aligned}
\hat{R}=g_{a b} R^{a b}\rightarrow1/(16 \pi(a^{2}-M^{2}+Q^{2})(2 a^{4}+4 M^{4}-5 M^{2} Q^{2}+Q^{4}+\\
4 M^{3} \sqrt{-a^{2}+M^{2}-Q^{2}}-3 M Q^{2} \sqrt{-a^{2}+M^{2}-Q^{2}}+a^{2}(-7 M^{2}+3 Q^{2}-5 M \sqrt{-a^{2}+M^{2}-Q^{2}}))^{5})
\end{aligned}
\end{equation}
According to the rewritten action quantity, we find that when the superradiance condition is established and the black hole does not undergo a thermodynamic phase transition, then the superradiance stability condition of the black hole is formed.

\section{Summary and discussion}
In this article, we introduced  $\mu=y\omega$\cite{38,39} into Kerr-Newman black holes and discussed the superradiant stability of Kerr-Newman black holes. We adopt the variable separation method to divide the motion equation of the minimally coupled scalar perturbation in Kerr-Newman black hole into two forms: angular and radial. 

The research\cite{40} in 2012 shows that the small mass asymptotically non-flat  Kerr-Newman-anti-de Sitter black holes produce superradiation instability under the scalar perturbation of charged mass. In 2016 \cite{41}, by analyzing the complex resonance spectrum of the charged mass scalar field in the near extremum  Kerr-Newman black holes space-time, the available dimensionless charge mass ratio is obtained $q/\mu$ to characterize the growth rate of super-radiative instability of scalar field, where $q$ is the charge of the scalar field. \cite{40} studied the small  Kerr-Newman-anti-de Sitter black hole, and \cite{41} studied the case without cosmological constant. The article\cite{10} studies the superradiation stability of Kerr-Newman black holes and charged scalar disturbances. They treat black holes as background geometry and study the equations of motion for scalar perturbations. Hod proved\cite{12} that the Kerr black hole should be superradiantly stable under massive scalar perturbation when $\mu \ge \sqrt{2}m\Omega_H$, where $\mu$ is the mass. In this article, a new variable y is added here to expand the results of the above article.
When $\sqrt{2(a^2+Q^2)}/{r^2_+}< \omega< m\varOmega_H+q\varPhi_H$,particularly $\mu \ge \sqrt{2}(m\varOmega_H+q\varPhi_H)$,
\begin{equation}
\begin{aligned}
16 \pi(a^{2}-M^{2}+Q^{2})(2 a^{4}+4 M^{4}-5 M^{2} Q^{2}+Q^{4}+\\
4 M^{3} \sqrt{-a^{2}+M^{2}-Q^{2}}-3 M Q^{2} \sqrt{-a^{2}+M^{2}-Q^{2}}+a^{2}(-7 M^{2}+3 Q^{2}-5 M \sqrt{-a^{2}+M^{2}-Q^{2}}))^{5}\neq0
\end{aligned}
\end{equation}
so the Kerr-Newman black hole is superradiantly stable at that time.

{\bf Data availability:}\\
 Derived data supporting the findings of this study are available from the corresponding author on request.

{\bf Acknowledgements:}\\
I would like to thank Jing-Yi Zhang in GuangZhou University and Yi-Xiao Zhang in South China Normal University for generous help.This work is partially supported by  National Natural Science Foundation of China(No. 11873025).

\end{document}